\documentstyle[epsfig,twoside,12pt]{article}
{\catcode `\@=11 \global\let\AddToReset=\@addtoreset}
\AddToReset{equation}{section}

\def\greaterthansquiggle{\raise.3ex\hbox{$>$\kern-.75em\lower1ex\hbox{$\sim$}}}
\def\lessthansquiggle{\raise.3ex\hbox{$<$\kern-.75em\lower1ex\hbox{$\sim$}}}

\newcommand{\beq}{\begin{equation}}
\newcommand{\eeq}{\end{equation}}
\newcommand{\beqa}{\begin{eqnarray}}
\newcommand{\eeqa}{\end{eqnarray}}
\newcommand{\beqan}{\begin{eqnarray*}}
\newcommand{\eeqan}{\end{eqnarray*}}
\newcommand{\ba}{\begin{array}}
\newcommand{\ea}{\end{array}}
\newcommand{\no}{\nonumber}

\newcommand{\Arccosh}{\rm Arccosh}
\newcommand{\sign}{\rm sign}

\newcommand{\ol}{\overline}
\newcommand{\ra}{\rightarrow}

\newcommand{\ve}{\varepsilon}
\newcommand{\vp}{\varphi}

\newcommand{\A}{{\cal A}}
\newcommand{\B}{{\cal B}}
\newcommand{\C}{{\cal C}}

\newcommand{\cS}{{\cal S}}

\newcommand{\V}{{\cal V}}

\newcommand{\dfrac}{\displaystyle \frac}

\def\nz{\ifmmode {I\hskip -3pt N} \else {\hbox {$I\hskip -3pt N$}}\fi}
\def\zz{\ifmmode {Z\hskip -4.8pt Z} \else
       {\hbox {$Z\hskip -4.8pt Z$}}\fi}
\def\qz{\ifmmode {Q\hskip -5.0pt\vrule height6.0pt depth 0pt
       \hskip 6pt} \else {\hbox
       {$Q\hskip -5.0pt\vrule height6.0pt depth 0pt\hskip 6pt$}}\fi}
\def\rz{\ifmmode {I\hskip -3pt R} \else {\hbox {$I\hskip -3pt R$}}\fi}
\def\cz{\ifmmode {C\hskip -4.8pt\vrule height5.8pt\hskip 6.3pt} \else
       {\hbox {$C\hskip -4.8pt\vrule height5.8pt\hskip 6.3pt$}}\fi}

\def\au{{\setbox0=\hbox{\lower1.36775ex%
\hbox{''}\kern-.05em}\dp0=.36775ex\hskip0pt\box0}}
\def\ao{{}\kern-.10em\hbox{``}}
\normalsize
\begin{document}
\bibliographystyle{plain}
\begin{titlepage}
\begin{flushright}
UWThPh--1999--22\\
ESI--695--1999\\
math-ph/9904027\\
%\today
April 26, 1999
\end{flushright}
\vspace*{3cm}   
\begin{center}
{\Large \bf  A pair potential supporting a mixed\\[10pt]
mean--field / BCS-- phase} %$^\star$}
\\[50pt]
N. Ilieva$^{\ast,\sharp}$ and W. Thirring  \\
\medskip
Institut f\"ur Theoretische Physik \\ Universit\"at Wien\\
%Boltzmanngasse 5, A-1090 Wien \\
\smallskip
and \\
\smallskip      
Erwin Schr\"odinger International Institute\\
for Mathematical Physics\\
\vfill
{\bf Abstract} \\
\end{center}
\vspace{0.4cm}
We construct a Hamiltonian which in a scaling limit becomes equivalent to
one that can be diagonalized by a Bogoliubov transformation. There may
appear simultaneously a mean-field and a superconducting phase. They
influence each other in a complicated way. For instance, an attractive
mean field may stimulate the superconducting phase and a repulsive one may
destroy it. 
\vfill 

{\footnotesize

%$^\star$ Work supported in part by ``Fonds zur F\"orderung der
%wissenschaftlichen Forschung in \"Osterreich" under grant P11287--PHY;

$^\ast$ On leave from Institute for Nuclear Research and Nuclear
Energy, Bulgarian Academy of Sciences, Boul.Tzarigradsko Chaussee 72, 1784
Sofia, Bulgaria

$^\sharp$ E--mail address: ilieva@ap.univie.ac.at}

\end{titlepage}
%\end{document}

\section*{Introduction}

In quantum mechanics a mean field theory means that the particle density
$\rho(x) = \psi^*(x)\psi(x)$ (in second quantization) tends to a {\it c}--number in
a suitable scaling limit. Of course, $\rho(x)$ is only an operator valued
distribution and the smeared densities $\rho_f = \int dx\,\rho(x)f(x)$ are
(at best) unbounded operators, so norm convergence is not possible. The
best one can hope for is strong resolvent convergence in a representation
where the macroscopic density is built in. The BCS--theory of
superconductivity is of a different type where pairs of creation operators
with opposite momentum $\tilde\psi^*(k)\,\tilde\psi^*(-k)$ ($\tilde\psi$
the Fourier transform and with the same provisio) tend to {\it
c}--numbers. This requires different types of correlations and one might
think that the two possibilities are mutually exclusive. We shall show
that this is not so by constructing a pair potential where both phenomena
occure simultaneously. On purpose we shall use only one type of fermions
as one might think that the spin--up electrons have one type of
correlation and the spin--down the other. Also the state which carries
both correlations is not an artificial construction but it is the
KMS--state of the corresponding Bogoliubov Hamiltonian. Whether the
phenomenon occurs or not depends on whether the emerging two coupled ``gap
equations" have a solution or not. This will happen to be the case in
certain regions of the parameter space (temperature, chemical potential, 
relative values of the two coupling constants). For simple forms of the
potentials these regimes will be explicitly shown. Our considerations hold
for arbitrary space dimension.

\section{Quadratic fluctuations in a KMS--state}

The solvability of the BCS--model \cite{BCS} rests upon the observation
\cite{H} that in an irreducible representation the space average of a
quasi--local quantity is a {\it c}--number and is equal to its ground
state expectation value. This allows one to replace the model Hamiltonian
by an equivalent approximating one \cite{NNB}. Remember that two
Hamiltonians are considered to be equivalent when they lead to the same
time evolution of the local observables \cite{TW}.

The same property holds on also in a temperature state (the KMS--state)
and under conditions to be specified later it makes the co--existence of
other types of phases possible.

To make this apparent, consider the approximating (Bogoliubov) Hamiltonian
\beqa
H_B' &=& \int dp\,\left\{\omega(p)a^*(p)a(p) +
\frac{1}{2}\Delta_B(p)\left[a^*(p)a^*(-p) + a(-p)a(p)\right]\right\}
\no \\
&=& \int W(p) b^*(p)b(p)\, ,
\eeqa
which has been diagonalized by means of a standard Bogoliubov
transformation with real coefficients (the irrelevant infinite constant in
$H_B'$ has been omitted)
\beqan
b(p) = c(p)a(p) + s(p)a^*(-p) \\
a(p) = c(p)b(p) - s(p)b^*(-p)
\eeqan
with
$$
c(p) = c(-p) \qquad s(p) = -s(-p)
$$
\beq
c^2(p) + s^2(p) = 1\, ,
\eeq
so that the following relations hold (keeping in mind that $\Delta, W,
s, c$ will be $\beta$--dependent)
$$
W(p) = \sqrt{\omega^2(p) + \Delta_B^2(p)} = W(-p) 
$$
\beq
c^2(p) - s^2(p) = \omega(p)/W(p)\,, \qquad
2c(p)s(p) = \Delta_B(p)/W(p)
\eeq

Hamiltonian (1.1) generates a well defined time evolution and a KMS--state
for the $b$--operators. For the original creation and annihilation
operators $a, a^*$ this gives the following evolution
$$
a(p) \ra a(p)\left(c^2(p)e^{-iW(p)t} + s^2(p)e^{iW(p)t}\right) - 
2ia^*(-p)c(p)s(p)\sin W(p)t
$$
and nonvanishing termal expectations 
\beqa
\langle a^*(p)a(p')\rangle &=&
\delta(p-p')\left\{\frac{c^2(p)}{1+e^{\beta(W(p)-\mu)}} +
\frac{s^2(p)}{1+e^{-\beta(W(p)-\mu)}}\right\} \no \\[2pt]
&:=& \delta(p-p')\{p\}\\ 
\langle a(p)a(-p')\rangle &=&
\delta(p-p')c(p)s(p)\tanh\frac{\beta(W(p)-\mu)}{2} :=
\delta(p-p')[p]
\eeqa
$$
\{p\} = \{-p\}, \qquad [p] = -[-p] 
$$
$c$ and $s$ are multiplication operators and are never Hilbert--Schmidt.
Thus different $c$ and $s$ lead to inequivalent representations and should
be considered as different phases of the system.

The expectation value of a biquadratic (in creation and annihilation
operators) quantity is expressed through (1.4,5)
\beqa
\langle a^*(q)a^*(q')a(p)a(p')\rangle =
\delta(q+q')\delta(p+p')[q][p]-& \no \\
- \delta(p-q)\delta(p'-q')\{p\}\{p'\} +
\delta(p-q')\delta(p'-q)\{p\}\{p'\} & 
\eeqa

So far we have written everything in terms of the operator valued
distributions $a(p)$. They can be easily converted into operators in the 
Hilbert space generated by the KMS--state by smearing with suitable test
functions. Thus, by smearing with e.g.
\beq
e^{-\kappa (p+p')^2-\kappa (q+q')^2}v(p)v(q), 
\qquad v \in L_2({\bf R}^d)
\eeq
one observes that in the limit $\kappa \ra \infty$ the first term in (1.6)
remains finite
$$
0 < \int dpdqv(p)v(q)[p][q] < \infty\, ,
$$
while the two others vanish
$$
\lim_{\kappa\ra\infty} \int dpdp' e^{-2\kappa(p+p')^2}v(p)v(p')\{p\}\{p'\}
= \lim_{\kappa\ra\infty} \kappa^{-3/2}\int dpv^2(p)\{p\}^2 = 0.
$$

Since we are in the situation of {\it Lemma 1} in \cite{IT}, we have thus
proved the following statement
\beq
\mbox{s-}\lim_{\kappa\ra\infty}\int
dpdp'\V(q,q',p,p')e^{-\kappa(p+p')^2}a(p)a(p') = \int
dp\V(q,q',p,-p)[p]
\eeq
for kernels $\V$ such that the integrals are finite.

With this observation in mind, a potential which acts for $\kappa
\ra \infty$ like (1.1) might be written as
\beq
V_B = \kappa^{3/2} \int
dpdp'dqdq'\,a^*(q)a^*(q')a(p)a(p')\V_B(q,q',p,p')\,
e^{-\kappa(p+p')^2-\kappa(q+q')^2}
\eeq
with $\V_B(q,q',p,p') = -\V_B(q',q,p,p')$ etc., in order to respect the
fermi nature of $a$'s. This potential has the property
$$
\begin{array}{ll}
\Vert V \Vert < \infty & \qquad \mbox{ for }\, \kappa <\infty \no \\[4pt]
\Vert V \Vert \ra \infty & \qquad \mbox{ for }\, \kappa \ra \infty \no
\end{array}
$$
Despite this divergence, potential (1.9) may still generate a
well--defined time evolution. The strong resolvent convergence in
(1.8) is essential, weak convergence would not be enough since it does 
not guarantee the automorphism property
$$
\tau_\kappa^t(ab) = \tau_\kappa^t(a)\tau_\kappa^t(b)\,\ra\,
\tau_\infty^t(ab) = \tau_\infty^t(a)\tau_\infty^t(b)\,.
$$
Note that the parameter $\kappa$
plays
in this construction the role of the volume from the considerations in
\cite{H}.

In the mean--field regime we want an effective Hamiltonian
\beq
H_B'' = \int dp \left[ \omega(p)a^*(p)a(p) +
\Delta_M(p)a^*(p)a(p)\right]\, .
\eeq
Here the KMS--state is defined for the operators $a, a^*$ themselves and
one should rather smear by means of 
\beq
e^{-\kappa(q-p)^2-\kappa(q'-p')^2}v(p)v(p')
\eeq
instead of (1.7), thus concluding that
\beq
\mbox{ s-}\lim_{\kappa\ra\infty} \int dpdq e^{-\kappa(q-p)^2} a^*(q)a(p)
\V_M(q,q',p,p') = -\int dp\frac{\V_M(p,q',p,p')}{1 +
e^{\beta(\ve(p)-\mu)}}\, ,
\eeq
with $\ve(p) = \omega(p) + \Delta_M(p)$. Relation (1.12) then suggests
another starting potential
\beq
V_M = \kappa^{3/2} \int
dpdp'dqdq'\,a^*(q)a^*(q')a(p)a(p')\V_M(q,q',p,p')\,
e^{-\kappa(q-p)^2-\kappa(q'-p')^2}
\eeq
with the same symmetry for the density $\V_M$ as in (1.9). However, in
both cases a Gaussian form factor in the smearing functions (1.7),(1.11)
has been chosen just for simplicity. In principle, this might be 
$C_o^\infty$ functions which have the $\delta$--function as a limit.

\section{The model}

Consider the following Hamiltonian
\beq
H = H_{kin} +  V_B +  V_M \, ,
\eeq
where $H_{kin}$ is the kinetic term and $ V_B, V_M$ are given by
(1.9),(1.13). The first potential term describes the superconducting
phase, similarly to the BCS--model, while the second corresponds to the
mean field regime. As already mentioned, both these terms diverge in the
limit $\kappa\ra\infty$. The solvability of the model for $\kappa
\ra \infty$ depends on whether or not it would be possible to replace
(2.1) by an equivalent Hamiltonian that might be readily diagonalized.
Remind that by equivalence of two Hamiltonians an equivalence of the time
evolution of the local observables they generate should be understood.
Therefore, the object of interest is the commutator of, say, a creation
operator with the potential. With (1.8), (1.12) taken into account, it
reads
\beq
[a(k), V] = 2\int dp\left\{c(p)s(p)\,[p]\,\V_B(k,-k,p,-p)a^*(-k) +
\V_M(p,k,p,k)\,\{p\}\,a(k)\right\}
\eeq
The Bogoliubov--type Hamiltonian for our problem should be a
combination of (1.1) and (1.10), that is of the form
\beq
H_B = \int dp \left\{ a^*(p)a(p)[\omega(p) + \Delta_M(p)] + 
\frac{1}{2}\Delta_B(p)[a^*(p)a^*(-p) + a(-p)a(p)]\right\}
\eeq
This Hamiltonian becomes equivalent to the model Hamiltonian (2.1),
provided the commutator $[a(k), H_B - H_{kin}]$ equals (2.2). Thus we are
led to a system of two coupled ``gap equations"
\beqa
\frac{1}{2} \Delta_M(k) &=& \int \V_M(k,p)\,
\left\{\frac{c^2(p)}{1 + e^{\beta(\ol W(p)-\mu)}} +
\frac{s^2(p)}{1+e^{-\beta(\ol W(p)-\mu)}}\right\}
\,dp \\[10pt]
\Delta_B(k) &=& \int \V_B(k,p)\, \frac{\Delta_B(p)}{\ol W(p)}
\,\tanh\frac{\beta(\ol W(p)-\mu)}{2}\,dp\, ,
\eeqa
with
\beq
\ol W(p) = \sqrt{[\omega(p) + \Delta_M(p)]^2 + \Delta_B^2(p)}\, .
\eeq
$c$ (and thus $s$, eq.(1.2)) are determined by either of the following
conditions 
\beq
c^2(p) - s^2(p) = [\omega(p)+\Delta_M(p)]/\ol W(p)\,, \qquad
2c(p)s(p) = \Delta_B(p)/\ol W(p)\, .  
\eeq 
The temperature and the interaction--strenght dependence of the system
(2.4--7) encode the solvability of the model. 

\section{Solution to the coupled gap equations}

For general potential densities $\V_M, \V_B$ solutions of the system
(2.4--6) cannot be explicitly written. In both low-- and high--temperature
limits a non--trivial ``mean--field gap" is possible, while for the
superconducting phase, characterized by a non--vanishing $\Delta_B$, a
critical temperature exists beyond which such a phase is no longer
possible.

However, for some simple though reasonable potentials solvability of
(2.4-7) and the behaviour of the solutions can be discussed in more
detail.

Thus, for the special case of an interaction concentrated about the Fermi
surface and being constant therein, potential densities can be chosen as
$$
\V_{B,M}({\bf k,p}) = \lambda_{B,M} \cS(\bf k) \cS(\bf p)
$$
with
$$
\cS({\bf k}) = \frac{1}{2\ve}\,
[\Theta(\vert{\bf k}\vert-\sqrt{\mu}+\ve) - \Theta(\vert{\bf
k}\vert-\sqrt{\mu}-\ve)],
$$
where $\Theta(x)$ is the Heaviside function. With the additional
assumption $\omega(p)=p^2$, the system (2.4--6) transforms for $\ve \ra 
0$ into
\beqa
\frac{1}{2} \Delta_M(\mu) &=&
\lambda_M\,\left\{\frac{c^2(\mu)}{1 +
e^{\beta(\ol W(\sqrt{\mu})-\mu)}} +
\frac{s^2(\mu)}{1+e^{-\beta(\ol W(\sqrt{\mu})-\mu)}}\right\} \\[6pt]
\ol W(\sqrt{\mu}) &=&  \lambda_B\,
\tanh\frac{\beta(\ol W(\sqrt{\mu})-\mu)}{2} \qquad \mbox{or }\qquad 
\Delta_B
= 0   
\eeqa
\beq
\ol W(\sqrt{\mu}) = \sqrt{[\mu + \Delta_M(\mu)]^2 + 
\Delta_B^2(\mu)}\\[4pt]
\eeq
with subsidiary conditions (2.7) correspondingly modified.

We shall always assume $\ol W(\sqrt{\mu})>0$. This is not really a
restriction, since the opposite situation might be similarly treated after
performing the exchange $b^* \leftrightarrow b$, also we might take
$\mu + \Delta_M \geq 0$. There are four energies $\lambda_M, \lambda_B,
\mu, \beta=1/T$ involved. The system exhibits severe dependence on their
relative values, in particular, the following holds: 
\begin{enumerate} 
\item[{\rm (i)}] for all values of $\lambda_M, \lambda_B, \mu$ there is a
purely mean-field solution
$$ 
\Delta_B = 0, \qquad \ol W(\sqrt{\mu}) - \mu = \Delta_M =
\frac{2\lambda_M}{1+e^{\beta \Delta_M}}\, , 
$$ 
but this will not be considered further; 
\item[{\rm (ii)}] $\ol W(\sqrt{\mu}) \leq \vert \lambda_B \vert$.
Furthermore, the condition $\ol W(\sqrt{\mu})>0$ implies 
$$ 
\lambda_B > 0\,\Longleftrightarrow \,\ol W(\sqrt{\mu}) > \mu, \qquad 
\lambda_B < 0\,\Longleftrightarrow \,\ol W(\sqrt{\mu}) < \mu; 
$$
\item[{\rm (iii)}] eq.(3.1) tells us that $\sign\,\Delta_M =
\sign\,\lambda_M\,$ and $\,\vert \Delta_M \vert \leq 2\vert \lambda_M
\vert\,$;
\item[{\rm (iv)}] %$\mu + \Delta_M \geq 0\,$, so the stability of the
%mean field regime is respected. Therefore 
$\ol W > \mu + \Delta_M\,$ and $\,\ol W \geq \Delta_B$. In addition, 
eq.(2.7) brings a restriction on the mixing angles in the Bogoliubov
transformation, $\vp \in [-\pi/4,\pi/4]\, \cup\,
[3\pi/4,-3\pi/4]$, so that $\sqrt{2}/2~\leq~\vert c(p)\vert \leq 1$.
\end{enumerate} 

Thus nontrivial solutions are only possible in the following regions:
\begin{enumerate}
\item[{\bf (a)}] $\lambda_B > 0: \quad \lambda_B > \mu\,$ \,\,(the area
$\A^+\cup \B^+\cup\C^+$ on Fig. 5); %\ol W(\mu)$;
\item[{\bf (b)}] $\lambda_B < 0: \quad -2\mu \leq \lambda_M \leq
-\dfrac{\mu T}{\vert \lambda_B \vert + 2T}$\\[4pt]
(limitations for the area $\A^-\cup\B^-$ on Fig. 5).
\end{enumerate}

Therefore two general cases have to be distinguished, corresponding to
attractive or repulsive superconducting potential, %that is to say
$\lambda_B<0$ or $\lambda_B>0$, that are qualitatively represented for
increasing values of $\vert \lambda_B \vert$ on Fig. 1.

%%%%%%%%%%%%%%%%%%%%%%%%%%%%%%%%%%%%%%%%%%%%%%%%%%%%%%%%%%%%%%%%%%%%%%%%%%
%%%%%%Fig.
\begin{figure}[h]
\setlength{\unitlength}{1pt}
\begin{picture}(490,155)
\put(25,0){\makebox(0,0)[lb]
{\scalebox{0.9}%
{\includegraphics*[0,235][490,410]{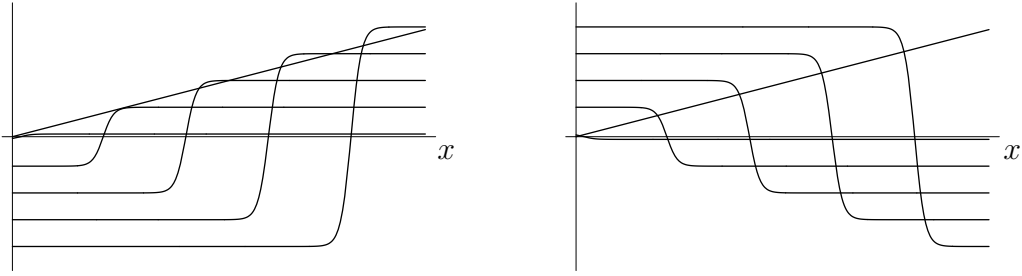}}}}
\put(200,48){\makebox(0,0)[l]{$x$}}
\put(414,48){\makebox(0,0)[l]{$x$}}
%\put(305,15){\makebox(0,0)[l]{{\small $\bar \lambda_B$}}}
\end{picture}
\caption{$\,\lambda_B\tanh{(x \mp k\lambda_B)}\,$ compared to the straight
line for increasing values of $\vert \lambda_B \vert$ % = 1,11,21,31,41$ 
and $k = 0.8$; $\,x = \beta \ol W(\sqrt{\mu})/2$: {\it(i)} $\,\lambda_B >
0$; {\it (ii)} $\,\lambda_B < 0$.}
\end{figure}
%%%%%%%%%%%%%%%%%%%%%%%%%%%%%%%%%%%%%%%%%%%%%%%%%%%%%%%%%%%%%%%%%

\vspace{0.6cm}

{\bf I.} \underline{$\lambda_B > 0$}

\vspace{6pt}

In this case necessarily only excitations with energies $> \mu$ may be
present. In both limits $\ol W \gg \mu + 2T$ and $\ol W \ll \mu + 2T$
solutions can be obtained analytically. 

\begin{itemize}
\item[{\bf I.A}]\, $0 < \ol W(\sqrt{\mu}) - \mu \gg 2T, \, \tanh{(x)} \ra
1$ 
\end{itemize}

In this regime one finds
\beqa
\ol W &=& \lambda_B \\
\Delta_M &=& \lambda_M \frac{\lambda_B - \mu}{\lambda_B + \lambda_M}\\
\Delta_B &=& \pm\frac{\lambda_B}{\lambda_B +
\lambda_M}\sqrt{(\lambda_B-\mu)(\lambda_B + \mu + 2\lambda_M)}
\eeqa
Thus a restriction has to be fulfilled
\beq
\lambda_B + \mu + 2\lambda_M > 0\, ,
\eeq
which is always the case by $\lambda_M > 0$ and also by negative
$\lambda_M$, if in addition 
$$
\vert \lambda_M \vert < (\lambda_B + \mu)/2\, ,
$$
that determines the area $\A^+\cup\B^+$ on Fig. 5. Then also the
mean--field gap energy becomes negative, though, as already mentioned, it
is possible to demand positivity for the ``pure" mean--field energy, $\mu
+ \Delta_M > 0$. 

\vspace{2pt}
\begin{itemize}
\item[{\bf I.B}]\, $0 < \ol W(\sqrt{\mu}) - \mu \ll 2T, \, \tanh{(x)} \ra
x $
\end{itemize} 
Here the solution reads
\beqa
\ol W(\mu) &=& \frac{\lambda_B\mu}{\lambda_B - 2T}\\[2pt]
\Delta_M &=& \lambda_M \\[2pt]
\Delta_B &=& \pm \sqrt{\frac{\lambda_B^2\mu^2}{(\lambda_B-2T)^2} -
(\mu+\lambda_M)^2}
\eeqa
and exists in the temperature interval
\beq
\frac{\lambda_M\lambda_B}{2(\mu+\lambda_M)} < T \ll
\frac{\lambda_B-\mu}{2}\, .
\eeq
The above restriction substanciates the idea of this limit as being valid
at sufficiently high, but nevertheless not at extremely high temperatures.

With an accuracy of $10^{-5}$ one can then estimate
$$
\lambda_M < \frac{\lambda_B - 4\mu}{4}\, ,
$$
so this asymptotic regime, when admissible, corresponds to the area
$\B^+\cup\C^+$ on Fig. 5. 

In the intermediate region an interesting phenomenon occurs that might be
visualized by the plot of the two sides of eq.(3.2) --- Fig. 2. There are
three different sets of values of the parameters of the theory, for which
one of the following situations is realised (see also Fig. 1(i)): 

%%%%%%%%%%%%%%%%%%%%%%%%%%%%%%%%%%%%%%%%%%%%%%%%%%%%%%%%%%%%%%%%%%%%%%%%%%
%%%%%%Fig.
\begin{figure}[h]
\setlength{\unitlength}{1pt}
\begin{picture}(470,150)
\put(0,0){\makebox(0,0)[lb]
{\scalebox{0.9}%
{\includegraphics*[0,0][470,150]{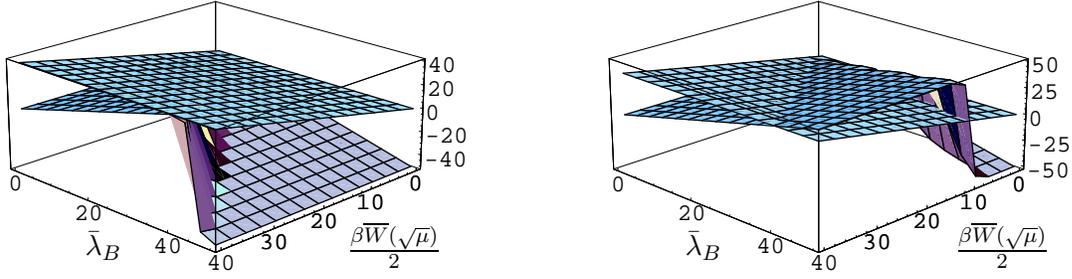}}}}
\put(140,15){\makebox(0,0)[l]{{\small $\frac{\beta \ol
W(\sqrt{\mu})}{2}$}}}
\put(41,15){\makebox(0,0)[l]{{\small $\bar \lambda_B$}}}
\put(370,15){\makebox(0,0)[l]{{\small $\frac{\beta \ol
W(\sqrt{\mu})}{2}$}}}
\put(268,15){\makebox(0,0)[l]{{\small $\bar \lambda_B$}}} 
\end{picture}
\caption{Plot of both sides of eq.(3.2) for $\lambda_B > 0\,$: {\it (i)}
$\, \mu=0.2\,\bar\lambda_B$; {\it (ii)} $\,\mu=0.9\,\bar\lambda_B$.}
\end{figure}
%%%%%%%%%%%%%%%%%%%%%%%%%%%%%%%%%%%%%%%%%%%%%%%%%%%%%%%%%%%%%%%%%%%%%%%%%%

\begin{enumerate}
\item[(a)] the co--existence of the mean--field regime and the
superconducting phase is not possible ($\Delta_M\not=0, \, \Delta_B=0$);
\item[(b)] for fixed values of the parameters such a
mixed phase is brought in to being and is uniquely determined
($\Delta_M\not=0,\,\Delta_B\not=0$);
\item[(c)] a kind of bifurcation occurs, so that two mixed phases
with different $\Delta_M$ and $\Delta_B$ become possible.
\end{enumerate}

%%%%%%%%%%%%%%%%%%%%%%%%%%%%%%%%%%%%%%%%%%%%%%%%%%%%%%%%%%%%%%%%%%%%%%%%%%
%%%%%%Fig.
\begin{figure}[ht]
\setlength{\unitlength}{1pt}
\begin{picture}(425,160)
\put(12,0){\makebox(0,0)[lb]
%{\scalebox{1}%
{\includegraphics*[5,220][430,380]{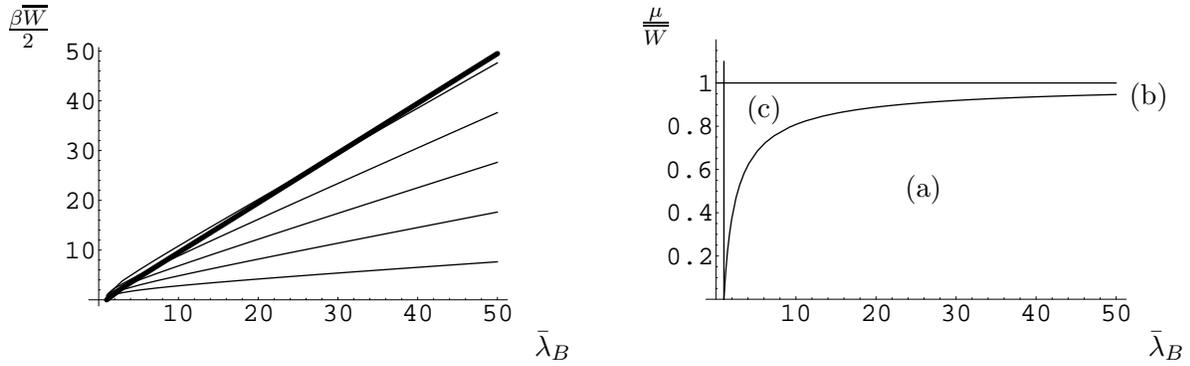}}}
\put(0,127){\makebox(0,0)[l]{{\small $\frac{\beta \ol W}{2}$}}}
\put(200,6){\makebox(0,0)[l]{{\small $\bar\lambda_B$}}}
\put(240,127){\makebox(0,0)[l]{{\small $\frac{\mu}{\ol W}$}}}
\put(432,6){\makebox(0,0)[l]{{\small $\bar\lambda_B$}}}
\put(340,65){\makebox(0,0)[l]{{\small (a)}}}
\put(425,100){\makebox(0,0)[l]{{\small (b)}}}
\put(280,95){\makebox(0,0)[l]{{\small (c)}}}
\end{picture}
\caption{{\it (i)\/}``Thermalized" energy $\beta\ol W(\sqrt{\mu})/2$ for
the equilibrium chemical potential, compared to the spectra for potentials
$\mu=k\,\lambda_B,\,\,k=0.1,0.3,0.5,0.7,0.9 \,\,$ ; {\it (ii)\/} Phase
diagram for the ratio $\mu/\ol W(\sqrt{\mu})$.}
\end{figure}
%%%%%%%%%%%%%%%%%%%%%%%%%%%%%%%%%%%%%%%%%%%%%%%%%%%%%%%%%%%%%%%%%%%%%%%%%%

More precisely, there exists a one--parameter family of equilibrium
chemical potentials, 
\beq
\bar\mu_e = -\Arccosh{\sqrt{\bar\lambda_B}} +
\bar\lambda_B\tanh{(\Arccosh{\sqrt{\bar\lambda_B}}\,)}\, ,
\eeq
$$
\bar Q := \beta Q/2\, ,
$$
for which solution of (3.1--3) is uniquely determined. The corresponding
energy is then found to be
$$
\frac{\beta \ol W(\sqrt{\mu})}{2} = \bar\lambda_B
\tanh{(\Arccosh{\sqrt{\bar\lambda_B}}\,)}\, .
$$
For chemical potentials $\bar\mu < \bar\mu_e$ there is no solution at all,
while for $\bar\mu > \bar\mu_e$ there are two solutions. The same is true
also for the ratio $\mu/\ol W(\sqrt{\mu})$ --- areas (a), (b) and (c) on
Fig. 3(ii) respectively. This, together with the comparison of the
``equilibrium" energy to the spectra for different allowed values of the
chemical potential --- Fig. 3(i), shows that for a given value of the
coupling constant $\lambda_B$, with the change of the temperature the
system can pass from one phase to another, so that a phase transition
occurs. When the critical value $T_c = \lambda_B/2$ is then reached,
the superconducting phase (either one, or two co--existing such phases) is
destroyed again. Furthermore, the two co--existing solutions correspond to
two different mixing angles, so are obtained through two different, hence
unequivalent, Bogoliubov transformations and this does not directly
afflict the local stability of the physical system. 

\vspace{0.6cm}

{\bf II.} \underline{$\lambda_B < 0$}

\vspace{6pt}

According to (3.2), for negative values of the coupling constant
$\,\lambda_B\,$ the excitations are necessarily with energies $\,\ol
W < \mu$. Thus, the mean--field gap must be negative, $\,\Delta_M < 0$,
that presupposes solvability of the model only when the second coupling
constant $\,\lambda_M\,$ is also negative. As one sees on the plot in Fig.
4, there is no nontrivial phase structure present in this regime and the
solution is always uniquely determined (see Fig. 1(ii)).

%%%%%%%%%%%%%%%%%%%%%%%%%%%%%%%%%%%%%%%%%%%%%%%%%%%%%%%%%%%%%%%%%%%%%%%%%%
%%%%%%Fig.
\begin{figure}[h]
\setlength{\unitlength}{1pt}
\begin{picture}(490,190)
\put(0,0){\makebox(0,0)[lb]
{\scalebox{0.9}%
{\includegraphics*[0,300][490,490]{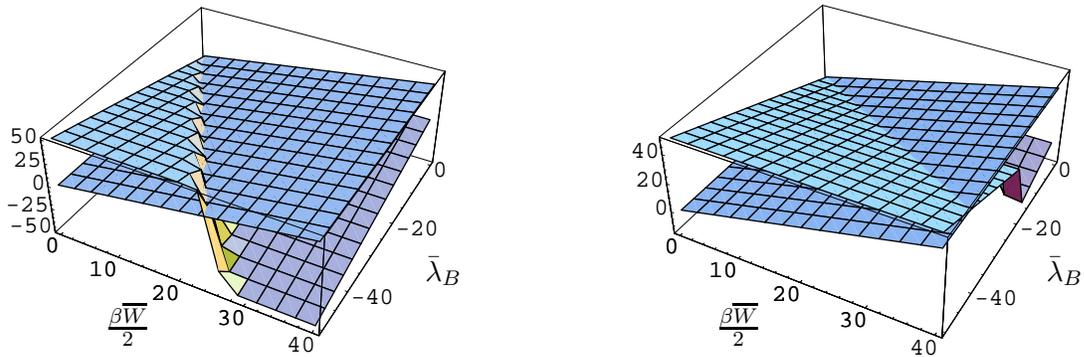}}}}
\put(49,35){\makebox(0,0)[l]{{\small $\frac{\beta \ol W}{2}$}}}
\put(170,55){\makebox(0,0)[l]{{\small $\bar\lambda_B$}}}
\put(280,35){\makebox(0,0)[l]{{\small $\frac{\beta \ol W}{2}$}}}
\put(405,55){\makebox(0,0)[l]{{\small $\bar\lambda_B$}}}
\end{picture}
\caption{Plot of both sides of eq.(3.2) for $\lambda_B < 0$: {\it
(i)} $\,\mu < \lambda_B\,$; {\it (ii)} $\,\mu > \lambda_B$.}
\end{figure}
%%%%%%%%%%%%%%%%%%%%%%%%%%%%%%%%%%%%%%%%%%%%%%%%%%%%%%%%%%%%%%%%%%%%%%%%%%

\vspace{2pt}
\begin{itemize}
\item[{\bf II.A}]\, $0 < \ol W(\sqrt{\mu}) \ll \mu - 2T $
\end{itemize}

The solution looks like the one for positive $\lambda_B$, eqs.(3.4--6),
only $\ol W = \vert \lambda_B \vert$. However, the relation between
the parameters changes
\beq
\lambda_B + \mu + 2\lambda_M < 0
\eeq
With the assumption $\mu + \Delta_M > 0$ this also means $ \lambda_B <
-\mu -\Delta_M $ but this does not strengthen the general restriction
{\bf (b)}.

\vspace{2pt}
\begin{itemize}
\item[{\bf II.B}]\, $0 < \ol W(\sqrt{\mu}) \gg \mu - 2T $
\end{itemize}

In this case the model is solvable, eqs.(3.8--10), within the temperature
interval
\beq
\frac{\mu + \lambda_B}{2} \ll T < \frac{\lambda_B\lambda_M)}{2(\mu +
\lambda_M)}\, 
\eeq 
and under the same assumption as in the case $\lambda_B > 0$, one
concludes that
$$
\lambda_M < -\frac{\lambda_B + 4\mu}{4} \qquad \mbox{ (area }\, \B^- \,
\mbox{on \, Fig.\,5)}
$$
The above restrictions for the allowed regions in the $(\lambda_B,
\lambda_M)$--plane, are depicted on Fig. 5. Nontrivial solutions are
possible in the area $\A^\pm\cup\B^\pm\cup\C^+$. In the $\A^\pm$--regions
in the parameter space only a highly excited system exhibits a
superconducting behaviour, while in the $\,\B^\pm$--regions this becomes
possible also for a system whose energy is close to the chemical
potential and in a restricted temperature interval. In both cases, when
all parameters fixed, solutions are uniquely defined.

%%%%%%%%%%%%%%%%%%%%%%%%%%%%%%%%%%%%%%%%%%%%%%%%%%%%%%%%%%%%% 
%%%%%%%%%% Fig. 
\begin{figure}[ht]
\setlength{\unitlength}{1pt}
\begin{picture}(320,210)
\put(60,0){\makebox(0,0)[lb] %{\scalebox{1}%
{\includegraphics*[0,340][320,550]{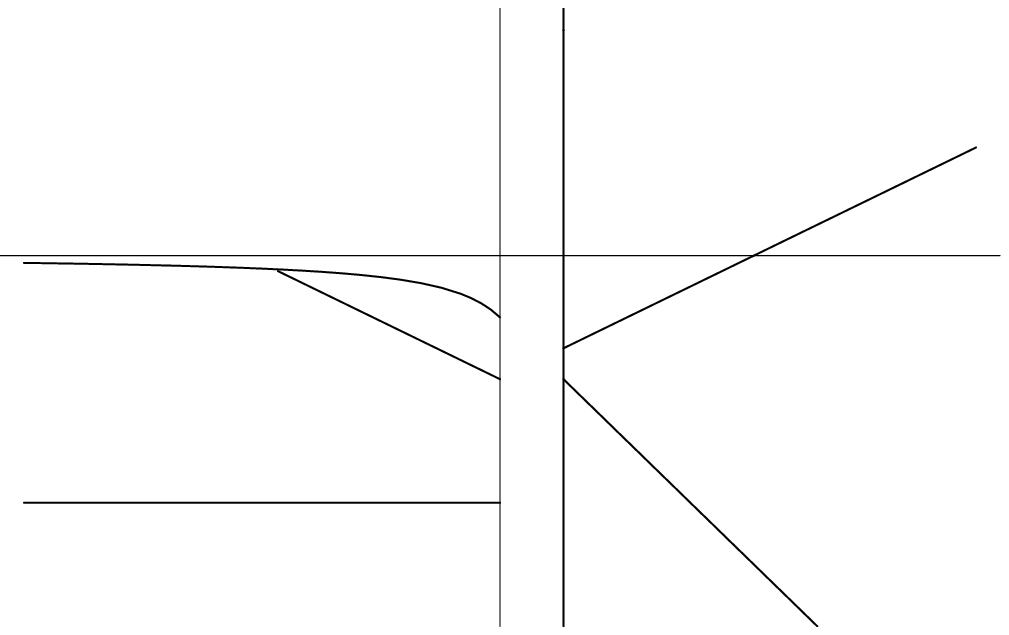}}}
\put(178,107){\makebox(0,0)[l]{{\small $\A^-$}}}
\put(140,90){\makebox(0,0)[l]{{\small $\B^-$}}}
\put(208,51){\makebox(0,0)[l]{{\scriptsize -$2\mu$}}}
\put(208,85){\makebox(0,0)[l]{{\footnotesize -$\mu$}}}
\put(208,103){\makebox(0,0)[l]{{\footnotesize -$\frac{\mu}{2}$}}}
\put(180,200){\makebox(0,0)[l]{{\small $\lambda_M$}}}
\put(225,129){\makebox(0,0)[l]{{\small $\mu$}}}
\put(262,156){\makebox(0,0)[l]{{\small $\A^+$}}}
\put(294,85){\makebox(0,0)[l]{{\small $\B^+$}}}
\put(250,30){\makebox(0,0)[l]{{\small $\C^+$}}}
\put(350,115){\makebox(0,0)[l]{{\small $\lambda_B$}}}
\end{picture} 
\caption{Allowed regions in the $(\lambda_B, \lambda_M)$--plane for given
$\mu$.}
\end{figure}
%%%%%%%%%%%%%%%%%%%%%%%%%%%%%%%%%%%%%%%%%%%%%%%%%%%%%%%%%%%%%%%%%%%%%%%%%%

\vspace{0.8cm}

\section{Conclusion}
Our model has four parameters, $\,\lambda_M, \lambda_B, \mu, T\,$, but by
scaling only their ratios are essential. For infinite temperature $\beta
= 0\,$  eqs.(3.1--3) admit only the mean field solution $\Delta_B = 0\,,\,
\Delta_M = \lambda_M,\,\, \ol W = \mu + \lambda_M$. This appears also
from Fig. 5: since the phase structure of the model and the very existence
of solutions depend on the ratios $\mu/T,\, \lambda_B/T,\, \lambda_M/T$, 
the area around the origin, so no superconducting phase present,
reflects the situation for $T \ra \infty$. By lowering the temperature one
meets also the BCS--type solution but in a rather complicated  region in
the 3--dimensional parameter space. 

However, the solutions themselves in the limiting cases {\bf I.A}, {\bf
II.A} are temperature independent while in {\bf I.B}, {\bf II.B} they do
depend on $T$. Therefore, for given values of $\lambda_M, \lambda_B$ and  
$\mu$  the type--{\bf A} solutions do not change upon heating
whereas the type--{\bf B} ones may appear and disappear, thus
bringing the system through regions (a), (b) and (c) on Fig. 3(ii).

Note that the limiting time evolution depends on the state. This would not
be the case if norm convergence would be present as shown in \cite{NT},
but we have only strong resolvent convergence at our disposal.  

Whenever $\lambda_B$ is positive, it must be also $ \,>\mu$.
Also for negative $\,\lambda_B,\, \lambda_M\,$ and $\,\lambda_M > -\mu\,$
there exists a finite gap for $\lambda_B$. A perturbation theory with
respect to $\lambda_B$ is in general doomed to failure since for no point
on the $\lambda_B = 0$ axis there is a neighbourhood full of the
$\Delta_B \not= 0$ phase.

It is interesting that without a mean field (the $\lambda_M = 0$ axis)
there are superconducting solutions only for $\lambda_B > \mu$. An
attractive mean field ($\lambda_M < 0$) stimulates superconductivity since
then it also appears for negative $\lambda_B$. However, too strong mean
field attraction destroyes it again. 

\section*{Acknowledgements} 
We are grateful to D.Ya. Petrina for stimulating our interest into the
problem and to N.N. Bogoliubov Jr. for pointing out to us the absence of
rigorous results to simultaneous mean--field -- BCS theory. We also
appreciate suggestive discussions with H. Narnhofer.

N.I. thanks the International Erwin Schr\"odinger Institute for
Mathematical Physics where the research has been performed, for
hospitality and financial support. This work has been supported
in part also by ``Fonds zur F\"orderung der wissenschaftlichen
Forschung in \"Osterreich" under grant P11287--PHY.

%\newpage


\begin{thebibliography}{99}
\bibitem{BCS} J. Bardeen, L.N. Cooper and J.R. Schrieffer, {\it Phys.
Rev.} {\bf 108 }, 1175 (1957).
\bibitem{H} R. Haag, {\it Nuovo Cim.} {\bf 25}, 287 (1962).
\bibitem{NNB} N.N. Bogoliubov, {\it Physica} {\bf 26}, 1 (1960).
\bibitem{TW} W. Thirring and A. Wehrl, {\it Commun. Math. Phys.} {\bf 4},
303 (1967).
\bibitem{IT} N. Ilieva, W. Thirring, {\it Eur. Phys. J.} {\bf C6}, 705 
(1999).
\bibitem{NT} H. Narnhofer, W. Thirring, {\it Phys. Rev. Lett.} {\bf 64},
1863 (1990).

\end{thebibliography}
\end{document}